\documentclass[10pt]{article}
\usepackage{amsmath,amssymb,tikz}
\usepackage{graphicx}
\usepackage{bm}
 \allowdisplaybreaks
 
 \newbox\pippobox
\newcounter{fig}   

\usepackage{upref}

\def \bo          {{\bm{\omega}}}

\def \bR          {{\bm{R}}}
\def\tri {\bigtriangledown}
\def\({\left(} \def\){\right)}
\def\[{\left[} \def\]{\right]}

\long\def\symbolfootnote[#1]#2{\begingroup%
\def\thefootnote{\fnsymbol{footnote}}\footnote[#1]{#2}\endgroup}

\newcommand{\aei}{\it Max Planck Institute for Gravitational Physics
(Albert Einstein Institute)\\ Am M\"uhlenberg 1, 14476 Golm,
Germany}

\newcommand{\itp}{\it Kavli Institute for Theoretical Physics,
Key Laboratory of Frontiers in Theoretical Physics,\\
Institute of Theoretical Physics, Chinese Academy of Sciences, Beijing 100190}

\newcommand{\yitp}{\it Yukawa Institute for Theoretical Physics (YITP),\\
Kyoto University, Kyoto 606-8502, Japan}
\begin{document}
\thispagestyle{empty}
\begin{center}

~\vspace{20pt}

{\Large\bf Entropy for gravitational Chern-Simons terms by squashed cone method}

\vspace{25pt}

Wu-Zhong Guo\symbolfootnote[1]{Email:~\sf wuzhong@itp.ac.cn}, Rong-Xin Miao\symbolfootnote[2]{Email:~\sf rong-xin.miao@aei.mpg.de}${}$

\vspace{10pt}${}^\ast{}$\itp

\vspace{10pt}${}^\ast{}$\yitp

\vspace{10pt}${}^\dagger{}$\aei

\vspace{2cm}

\begin{abstract}
In this paper we investigate the entropy of gravitational Chern-Simons terms for the horizon with non-vanishing extrinsic curvatures, or the holographic entanglement
entropy for arbitrary entangling surface. In 3D we find no anomaly of entropy appears. But the squashed cone method can not be used directly to get the correct
result. For higher dimensions the anomaly of entropy would appear, still, we can not use the squashed cone method directly. That is becasuse the Chern-Simons action is not gauge invariant. To get a reasonable result we suggest two methods. One is by adding a boundary term to recover the gauge invariance. This boundary term can be derived from the variation of the Chern-Simons action. The other one is by using the Chern-Simons relation $d\bm{\Omega_{4n-1}}=tr(\bm{R}^{2n})$. We notice that the entropy of $tr(\bm{R}^{2n})$ is a total derivative locally, i.e. $S=d s_{CS}$. We propose to identify $s_{CS}$ with the entropy of gravitational Chern-Simons terms $\Omega_{4n-1}$. In the first method we could get the correct result for Wald entropy in arbitrary dimension. In the second approach, in addition to Wald entropy, we can also obtain the anomaly of entropy with non-zero extrinsic curvatures. Our results imply that the entropy of a topological invariant, such as the Pontryagin term $tr(\bm{R}^{2n})$ and the Euler density, is a topological invariant on the entangling surface.
\end{abstract}

\end{center}

 \newpage

\tableofcontents

\section{Introduction}
The entropy is often used as a quality to reflect the degree of freedoms of a system. In the gravitational field black hole entropy, i.e., Bekenstein-Hawking entropy\cite{Bekenstein}\cite{Hawking}, is related with
geometry of the spacetime, and performs as a thermal quality. On the other hand in the gauge field theory with gravity dual, the entanglement
entropy for a subsystem could also have a geometry description in the gravity side, which is known as the Ryu-Takayanagi formula\cite{Ryu:2006bv}. In general the geometry description of the entropy is closely connected with the detail of the theory. Wald formula \cite{Wald}provides the connection between
the action and entropy for general covariant gravitational theory. \\
The recent idea concerning about the generalized gravitational entropy \cite{Lewkowycz:2013nqa} gives a strong evidence for the Ryu-Takayanagi formula on general entangling surfaces. Generalization to theory other than Einstein gravity seems not so straightforward. To deal with  the singular mainfold without a $U(1)$ symmetry in the subspace orthogonal to singular surface $\Sigma$, one must also consider the possible contribution from the extrinsic curvatures. The paper \cite{Fursaev:2013fta} provides a method to calculate the geometry quality of this kind singularity, called by the squashed cone method. More systemic study on this problem for different covariant gravity theory can be found in \cite{Dong:2013qoa,Camps:2013zua}, see also\cite{Miao0}-\cite{Bhattacharyya:2014yga}.

The squashed cone method works well for covariant gravitational theories. However, it may break down or need to be modified for non-covariant gravitational theories. There are two kind of non-covariant gravitational theories. The first one is that neither the action or the equations of motion are gauge invariant. The balck hole thermodynamics is not well-defined for this kind of gravitational theories \cite{Dubovsky,Jacobson,Miao3}. The second one is that the action is gauge invariant up to some boundary terms.  Theory with gravitational Chern-Simons term is of this kind of non-covariant theories. The Chern-Simons(CS) term as
a possible correction to Einstein gravity is motivated by the low-energy effective action from superstring theories. In 3D the modification of
the black hole entropy by CS term is studied in many literatures, see \cite{Solodukhin:2005ah}-\cite{Park:2006gt}. In higher dimensions the answer is also found in \cite{Bonora:2011gz}\cite{Bonora:2012xv} by generalizing the covariant phase formalism. There are also some studies on the contribution to thermodynamics and transport in hydrodynamics from the gravitational anomalies\cite{Jensen:2013kka}\cite{Jensen:2013rga}, which is related to the gravitational CS term by AdS/CFT\cite{Azeyanagi:2013xea}\cite{Azeyanagi:2014sna}\cite{Azeyanagi:2015gqa}.
In this paper we would like to study the problem for generalized gravitational entropy. Specially, we would use the regularization process developed in \cite{Dong:2013qoa}. The method works well for covariant theory, but for CS theory, we find that the method should be
modified significantly to get a consistent result. The modification is related with the local gauge transformation of theory with gravitational Chern-Simons term, for this transformation would produce a total derivative terms. Actually the regularized process would ignore the possible effect caused by the gauge transformation. The modification is based on the consideration to fix the gauge freedom. In the following we refer to entropy either to generalized gravity entropy or to holographic entanglement entropy for they are the same thing in a sense.\\

The paper is organized as follows. In the next section we briefly introduce the regularization process developed in \cite{Dong:2013qoa}.  In section 3 we briefly review the gravitational Chern-Simons terms in arbitrary dimension. We will calculate the entropy in 3D in this section 4. In section 5, we propose an approach to derive the entropy of gravitational Chern-Simons terms. We work out the entropy exactly in 7D space-time. In section 6, we use this method to get the Wald entropy in arbitrary dimension. We will also discuss other approach
to get the correct Wald entropy. We will conclude and discuss some related problems in section 7. Some useful formula and detail calculation can be found
in Appendix.

\section{Review of Generalized Gravitational Entropy}
The generalized gravitational entropy is based on the ``replica trick'' in Euclidean spacetime. In classical approximation the density matrix $\rho $ of the gravity field would be related with the Euclidean solution by $tr\rho =I$, where $I$ is the on-shell Euclidean action. The n-th replica spacetime $B_n$ would produce the relation $tr \rho^n=I(n)$. One could consider the orbifold $\hat B_n\equiv B_n/Z_n$. This leads to $I_n=n\hat{I}(n)$, where $\hat{I}(n)$ means the action with the solution $\hat B_n$ without counting the contribution from the conical defect in $\hat B_n$ . The entropy in $O(G^{-1})$ can be expressed as
\begin{eqnarray}
S=\partial_\epsilon \hat{I}(\epsilon),
\end{eqnarray}
where we denote $n=1+\epsilon$. Now one fills the singular cone, the calculation becomes
\begin{eqnarray}\label{Keyformula}
S=-\partial_\epsilon I(\epsilon),
\end{eqnarray}
where $I(\epsilon)$ is the action of the regularized squashed cone. This equation is the starting point to calculate the entropy. We refer the readers to \cite{Lewkowycz:2013nqa}and \cite{Dong:2013qoa} for more argument and explanation to the generalized gravitational entropy and holographic entanglement entropy.\\
But in the theory that is not-covariant this statement can not be used directly. Such as the gravity with Chern-Simons term we will discuss below, the local gauge
transformation or the non-covariant part of the diffeomorphism will lead to a boundary term, which also contributes to the $O(\epsilon)$. Our main discussion below is about how to eliminate the ambiguity in the non-covariant theory when using the squashed cone method.\\

Regardless of the difference that we mention above, one have to find a way to regularize the squashed cone . We would follow the regularization process in \cite{Dong:2013qoa}. According to \cite{Dong:2013qoa}, the metric of regularized cone is
\begin{eqnarray}\label{metric}
ds^2&=&e^{2A}[dzd\bar{z}+e^{2A}T(\bar{z}dz-zd\bar{z})^2]+\big{(} g_{ij}+2K_{aij}x^a+Q_{abij}x^ax^b\big{)}dy^idy^j\nonumber\\
&+&2i e^{2A}(U_i+V_{ai}x^a)(\bar{z}dz-zd\bar{z})dy^i+...
\end{eqnarray}
where  $T,g_{ij}, K_{aij},Q_{abij},U_i,V_{ai}$ are independent of z and $\bar{z}$, with the exception that $Q_{z\bar{z}ij}=Q_{\bar{z}zij}$ contains a factor $e^{2A}$. The warp factor $A$ is regularized by a thickness parameter $a$ as $A=-\frac{\epsilon}{2}\log(z\bar{z}+a^2)$. The result is independent of the choice of regularization.

The contribution from the Wald entropy is related with the fact
\begin{eqnarray}\label{Waldcontribution}
\int dz d\bar ze^{-\beta A} \partial_z \partial_{\bar z}A =-\pi \epsilon.
\end{eqnarray}
The key observation of \cite{Dong:2013qoa} is that
\begin{eqnarray}\label{Key}
\int \rho d\rho \partial_z A\partial_{\bar{z}} A e^{-\beta A}=-\frac{\epsilon}{4\beta},
\end{eqnarray}
where $z=\rho e^{i\tau}$. The would-be logarithmic divergence gains a $\frac{1}{\epsilon}$ enhancement:
\begin{eqnarray}\label{Key1}
\int \rho d\rho \frac{1}{\rho^2} e^{\beta\epsilon } \sim \frac{1}{\beta \epsilon}.
\end{eqnarray}
This will give the anomaly contribution of the entropy. One is suggested to refer the recent paper \cite{Miao0} in which we discuss more possible terms
that may contribute to entropy. For our purpose in this paper (\ref{Waldcontribution})(\ref{Key}) are enough.

\section{Gravitational Chern-Simons term}
In the this section we would like to introduce some definitions and properties of CS terms.
The gravitational CS terms can be constructed in two different ways, one is by the one-form of Christoffel symbol $\mathbf{\Gamma}$, the other one is the spin connection $\bm{\omega}$.\\
By using $\bm{\omega}$ the (2n+1)-dimensional gravitational
CS terms $\bm{\Omega_{2n+1}}$ are formally defined as
\begin{eqnarray}\label{Chern-Simons}
d\Omega_{2n+1}(\bm{\omega})=Tr\mathbf{R}^{n+1},
\end{eqnarray}
where $\mathbf{R}=d\bm{\omega}+\bm{\omega}\bm{\omega}$ is the two-form curvature, and we suppress the wedge between the forms.
$\bm{\Omega_{2n+1}}$ can be expressed as
\begin{eqnarray}\label{Chern-Simons terms}
\bm{\Omega_{2n+1}}=(n+1)\int_0^1t^n str(\bm{\omega}\bm{R_t}^n),
\end{eqnarray}
where $\bm{R_t}^n\equiv \bm{R} +(t-1)\bm{\omega}^2 $, and
``str'' is defined by
\begin{eqnarray}
str(A_1,A_2,...,A_n) \equiv \frac{1}{n!}\sum_{\pi} Tr(A_{\pi(1)}A_{\pi_(2)}...A_{\pi(n)}),
\end{eqnarray}
$\pi$ denotes the permutations of \{1,2,...,n\}.
 The CS action is
\begin{eqnarray}\label{CSaction}
I_{CS}=\frac{\lambda}{32\pi G}\int_{M_{2n+1}}\frac{2^n}{n+1}\bm{\Omega_{2n+1}},
\end{eqnarray}
$\lambda$ is the coupling constant. The full action is
\begin{eqnarray}
I=\frac{1}{16\pi G}\int d^{2n+1}x \sqrt{-g}(R+\frac{n(2n+1)}{l^2})+I_{CS}.
\end{eqnarray}\\
The spin connection $\bm{\omega}$ can be construct by vielbeins $E=\{e^{a_{\nu}}_{\mu}\}$, which is defined by $G_{\mu \nu}=e^{a_{\kappa}}_{\nu}e^{a_{\sigma}}_{\mu}\delta_{a_{\kappa}a_{\sigma}}$.
As an example we could choose the vielbeins of (\ref{metric}) up to $O(\rho)$ as follows.
 \begin{eqnarray}\label{vielbein}
&&e_{ a_1}=e_{\mu a_1}dx^{\mu}=\frac{e^A}{2}(dz+d\bar{z})+e^A(\bar{z}-z) U_idy^i,\nonumber\\
&&e_{ a_2}=-i\frac{e^A}{2}(dz- d\bar{z})-e^{A}(\bar{z}+z) U_idy^i,\nonumber\\
&&e_{a_i}=\bar{e}_{ja_i}dy^j+x^aK_{aja_i}dy^j,
\end{eqnarray}
where $\bar{e}_{ja_i}\bar{e}_{ka_i}=g_{jk}$ and $K_{aja_i}=K_{aji}e_{a_i}^{i}$. Here $a_1,a_2$ denotes the local Lorentz indices with respect to $z,\bar{z}$. One can check that the above vielbeins can yield the correct metric in order $O(\rho)$:
\begin{eqnarray}\label{vmetric}
e_{\mu a_1}e_{\nu a_1}+e_{\mu a_2}e_{\nu a_2}+e_{\mu a_i}e_{\nu a_i}=G_{\mu\nu}+O(\rho^2).
\end{eqnarray}
The choose of vielbeins are not unique, different ones are
related by performing local Lorentz transformations,
\begin{eqnarray}\label{Lorentz transformation}
e'_{a_{\mu}}=\Lambda(x)_{a_{\mu}}^{\ b_{\nu}}e_{b_{\nu}}.
\end{eqnarray}
We can calculate the spin connection $\bm{\omega}$ by the following formula,
\begin{eqnarray}\label{Spincalculation}
\omega_{\mu,a_\nu a_\kappa}&=&\frac{1}{2}(C_{{a_\nu} \sigma \mu}e^\sigma_{a_\kappa}+ C_{a_\kappa \mu\sigma}e^\sigma_{a_\nu}-C_{a_\sigma \alpha \beta }e^\alpha_{a_\nu}e^\beta_{a_\kappa}e^{a_{\sigma}}_\mu ) \nonumber \\
C^{a_\sigma}_{\mu \nu}&=&\partial_\mu e ^{a_\sigma}_\nu-\partial_\nu e ^{a_\sigma}_\mu.
\end{eqnarray}
The spin connection $\bm{\omega}$ can also be related with the one-form Christoffel symbols $\mathbf{\Gamma}$ by the vielbeins,
\begin{eqnarray}\label{spinc}
\omega_{\mu\ \ y}^{\ x}=\Gamma^{x}_{\mu y}-e^{ax}\partial_{\mu}e_{y a},
\end{eqnarray}
 where we have mapped all the index into spacetime. We list some useful components of the Christoffel symbols and spin connections in Appendix A.\\

By varying the action (\ref{Chern-Simons terms}) new terms would contribute to the equation of motion. The result is derived in \cite{Bonora:2011gz}\cite{Solodukhin:2005ns},
\begin{eqnarray}
R_{\mu \nu}-\frac{1}{2}G_{\mu \nu}R-\frac{n(2n+1)}{l^2}G_{\mu\nu}+ \lambda C_{\mu\nu}=0
\end{eqnarray}
where
\begin{eqnarray}
C^{\mu\nu}=\bigtriangledown_\alpha S^{\mu\nu \alpha}
\end{eqnarray}
with
\begin{eqnarray}
S^{\mu\nu \alpha}=-\frac{1}{2}\epsilon^{\lambda_1 \lambda_2...\lambda_{2n}\mu}R^{\nu}_{\ \kappa_1\lambda_1\lambda_2}
R^{\kappa_1}_{\ \kappa_2\lambda_3\lambda_4}...R^{\kappa_{2n-2}\alpha}_{\ \ \ \lambda_{2n-1}\lambda_{2n}}.
\end{eqnarray}
The CS term expressed by $\mathbf{\Gamma}$ is similar to $\bm{\omega}$, with $\bm{\omega}$ replaced by $\mathbf{\Gamma}$ in (\ref{Chern-Simons terms}).

\section{Entropy of gravitational Chern-Simons terms in 3D}
In the case of 3D theory with CS  term is also known as topologically massive gravity\cite{DJT2}. The black hole entropy in this theory has been discussed in many authors, see e.g. \cite{Solodukhin:2005ah}-\cite{Tachikawa:2006sz} . Non-trivial correction appears for rotating BTZ black. We would like to use the squashed cone method to study the correction of CS term to black hole entropy (also the HEE formula) in 3D spacetime. \\
For the Euclidean theory, the action of 3D CS term  is
\begin{eqnarray}\label{3dCSGamma}
I_{CS}=\frac{-i\lambda}{64\pi G}\int d^3x \sqrt{G}\epsilon^{\mu \nu \sigma}(\Gamma_{\beta \mu}^{\alpha}R^\beta_{\ \alpha \nu \sigma}-\frac{2}{3}\Gamma_{\beta \mu}^{\alpha}\Gamma_{\gamma \nu }^{\beta}\Gamma_{\alpha \sigma}^{\gamma}),
\end{eqnarray}
where we have integrated $t$, $-i$ appears because of the Wick rotation, see \cite{Bonora:2012xv}.\\
\subsection{Result by using the squashed cone method directly}
 Firstly we use the squashed cone method directly. To get the contribution to HEE from the CS action, we need work in metric (\ref{metric}), and find the $O(\epsilon)$. We calculate the components of $\bm{\Gamma}$ and $\bm{R}$ in the Appendix A. It's easy to see that only the Wald
entropy appears, and the final result of the entropy is
\begin{eqnarray}\label{3DEntropyGamma}
S=\frac{ i\lambda}{16G} \int_\Sigma dy (\Gamma_{yz}^{z}\epsilon^{ z}_{\  z}+\Gamma_{y\bar z}^{\bar z}\epsilon^{\bar z}_{\ \bar z})= \frac{i \lambda}{16G} \int_\Sigma  \Gamma_{\sigma \mu}^{\nu}\epsilon_{\ \nu}^{\mu}dx^{\sigma}
\end{eqnarray}
where $\Sigma$ is codimension-2 surface in the bulk. \\
It's also interesting to check the result by using the spin connection $\bm{\omega}$. The action would be
\begin{eqnarray}\label{3dCSomega}
I_{CS}=\frac{-i \lambda}{64\pi G}\int d^3x \sqrt{G}\epsilon^{\mu \nu \sigma}(\omega_{\mu\ \beta}^{\ \alpha}R^\beta_{\ \alpha \nu \sigma}-\frac{2}{3}\omega_{\mu\ \beta}^{\ \alpha}\omega_{\nu\ \gamma }^{\ \beta}\omega_{\sigma\ \alpha}^{\ \gamma}).
\end{eqnarray}
We also calculate all the components of $\bm{\omega}$ in appendix A. Besides the Wald entropy there are also contributions from the second term in (\ref{3dCSomega}). But the contribution finally vanishes. The result is also
\begin{eqnarray}\label{3dEntropyomega}
S=\frac{ i\lambda}{16G} \int_\Sigma  \Gamma_{\sigma \mu}^{\nu}\epsilon_{\ \nu}^{\mu}dx^{\sigma},
\end{eqnarray}
where we use $\omega_{y\ z}^{\ z}=\Gamma_{yz}^z+O(z)$. It is natural because $\bm{\Gamma}$ ia related with $\bm{\omega}$ by local gauge transformation, and the result is gauge-invariant.\\
But (\ref{3DEntropyGamma})(\ref{3dEntropyomega}) is not consistent with the result given in literature \cite{Solodukhin:2005ah}-\cite{Park:2006gt}.  The inconsistence warn us to be careful when dealing with the non-covariant theory.
Just like the Wald method \cite{Wald} to calculate the entropy of CS term, some modification is expected to get the correct result, see \cite{Tachikawa:2006sz}. We would give a solution to this problem in the next subsection for the squashed cone method.

\subsection{A solution}

The vielbeins (\ref{vielbein}) still have a gauge freedom (\ref{Lorentz transformation}). An arbitrary local Lorentz transformation would
produce an additional total derivative term for the action. This term contributes to entropy if we use the squashed cone method before integrating out the total derivation term. The freedom should be eliminated if we want a reasonable result.\\
We denote the action of CS term after regularization as $I(\epsilon)$, the entropy $S=-\partial_{\epsilon}I(\epsilon)$. As we can see from
metric (\ref{metric}) the vielbeins $E^a_\mu$ and the spin connection $\omega_{\mu\ b}^{\ a}$ would also depend on $\epsilon$.
Under an infinitesimal local Lorentz transformation parameterized by $\theta_{\ b}^a$,
\begin{eqnarray}\label{Transformationveilbern}
&&\delta_\theta e^b=-\theta_{\ a}^b  e^a,\nonumber \\
&&\delta_\theta \bm{\omega}_{ \ b}^{\ a}=d \theta^a_{\ b} +[\bm{\omega},\theta]^a_{\ b}.
\end{eqnarray}
The 3D CS action
\begin{eqnarray}
I^{(3)}_{CS}=\frac{-i\lambda}{32\pi G}\int_{M_3} Tr(\bm{\omega}d\bm{\omega}+\frac{2}{3}\bm{\omega}^3)
\end{eqnarray}
would have a variation that is total derivative term,
\begin{eqnarray}\label{Variance3Daction}
\delta_\theta I^{(3)}_{CS} =\frac{-i\lambda}{32\pi G} \int_{M_3} d Tr(\bm{\theta} d\bm{\omega}(\epsilon)).
\end{eqnarray}
When we work in the metric (\ref{metric}), the above term (\ref{Variance3Daction}) contains $\epsilon$, which contributes to entropy.
This suggests new terms should be added to the action to eliminate the ambiguity. We find the following one satisfies the
requirement,
\begin{eqnarray}\label{Addtermfor3DCS}
\Delta I^{(3)}_{CS} = \frac{i\lambda}{32\pi G} \int_{M_3} d Tr( \bm{\omega}(0)\bm{\omega}(\epsilon)).
\end{eqnarray}
The total derivative term does not modify the equation of motion. At the limit $\epsilon\to 0$ it vanishes, we get the original
action. Under the infinitesimal local Lorentz transformation,
\begin{eqnarray}
\delta_\theta \Delta I^{(3)}_{CS} &=& \frac{i\lambda}{32\pi G} \int_{M_3} d Tr( \delta_\theta \bm{\omega}(0)\bm{\omega}(\epsilon)+\bm{\omega}(0)\delta_\theta \bm{\omega}(\epsilon))\nonumber \\
&=& \frac{i\lambda}{32\pi G} \int_{M_3} \Big(d Tr\bm{\theta} d \bm{\omega}(\epsilon)-d Tr\bm{\theta} d \bm{\omega}(0)\Big)
\end{eqnarray}
The variation of the modified action,
\begin{eqnarray}
\delta_\theta  \tilde{I}^{(3)}_{CS}= \delta_\theta (I^{(3)}_{CS}+\Delta I^{(3)}_{CS})=\frac{i\lambda}{32\pi G} \int_{M_3}d Tr\bm{\theta} d \bm{\omega}(0).
\end{eqnarray}
This is what we expected. Now the contribution from (\ref{Addtermfor3DCS}) is
\begin{eqnarray}
\Delta S=\frac{i \lambda}{16G} \int_\Sigma  \Gamma_{\sigma \mu}^{\nu}\epsilon_{\ \nu}^{\mu}dx^{\sigma}.
\end{eqnarray}
Including this contribution we recover previous result for black hole in 3D gravity with CS term. Then the HEE formula for 3D gravity theory with CS term is
\begin{eqnarray}\label{3DCSformula}
\frac{1}{4G}\int dy \sqrt{g_{yy}}+ \frac{i\lambda}{8G}\int \Gamma_{\sigma \mu}^{\nu}\epsilon^\mu_{\ \nu}dx^{\sigma}.
\end{eqnarray}\\
In \cite{Solodukhin:2005ah} and \cite{Castro:2014tta} the authors actually use different method to get the correct result. As the statement of \cite{Solodukhin:2005ah}, $\omega=\omega_{sing}+\omega_{reg}$, so $R=d\omega_{sing}+d\omega_{reg}+...$, $\int\omega R =\int \omega_{reg} R_{sing}+\omega_{sing}d\omega_{reg}+...=2\int \omega_{reg} R_{sing}...$. A factor $2$ also appears in this approach.  In higher dimension the anomaly term of the entropy (\ref{Key}) will appear, it seems very difficult
to perform the similar process to get the correct result. We consider that the ambiguity of result is related with the local gauge transformation.
To eliminate the ambiguity of the gauge transformation, one need to add a boundary term, which will not effect the equation of motion.  But is
also very hard to construct the suitable boundary term like (\ref{Addtermfor3DCS}) in more general case. We will comment on the problem in higher dimension theory with CS term later.

\subsection{The surface}

If we know the solution of a black hole in 3D, the result (\ref{3DEntropyGamma}) can be directly used to calculate the entropy. But for HEE we have to find the surface $\Sigma$ firstly. We would use boundary condition method to determine where the $\Sigma$ should be. We will follow the same strategy as \cite{Lewkowycz:2013nqa}\cite{Dong:2013qoa}.\\
We could parameterize the coordinate $y$ in 3D, the metric is
\begin{eqnarray}
ds^2=e^{2A}[dzd\bar{z}+e^{2A}T(\bar{z}dz-zd\bar{z})^2]+\big( 1+2K_{a}x^a+Q_{ab}x^ax^b\big)dy^2\nonumber\\
+2i e^{2A}(U+V_{a}x^a)(\bar{z}dz-zd\bar{z})dy+...
\end{eqnarray}
In three dimensions
\begin{eqnarray}
C^{\mu\nu}=\epsilon^{\mu \kappa \sigma}\bigtriangledown_\kappa (R^\nu _\sigma-\frac{1}{4}\delta^\nu _\sigma R)
\end{eqnarray}
The equation of motion is
\begin{eqnarray}
E_{\mu\nu}=R_{\mu \nu}-\frac{1}{2}G_{\mu \nu}R-\frac{1}{l^2}G_{\mu\nu}-i\lambda C_{\mu\nu}=0
\end{eqnarray}
Let's check the divergent terms in $E_{zz}$. The result is
\begin{eqnarray}\label{3Ddivergentterm}
E_{zz}=\frac{\epsilon}{z}\Big(-\frac{1}{2}K_z(y)+\lambda (U(y)-3V_z)K_z(y)-\frac{i\lambda}{2} K'_{z}(y)\Big)+...,
\end{eqnarray}
where the ``...'' means terms less divergent. We should set the divergent term in (\ref{3Ddivergentterm}) to zero. We get the constraint,
\begin{eqnarray}\label{3dConditionK}
-\frac{1}{2}K_z(y)+\lambda (U(y)-3V_z)K_z(y)-\frac{i\lambda}{2} K'_{z}(y)=0,
\end{eqnarray}
when $\lambda=0$ we get the conditions for Einstein gravity (\cite{Lewkowycz:2013nqa}). For $E_{\bar z \bar z}$ we would get a constraint on $K_{\bar z}$ with $z\leftrightarrow \bar z$ and $U(y) \leftrightarrow -U(y))$, $V_{z}\to -V_{\bar z}$ in (\ref{3dConditionK}). The CS term would give a non-trivial correction to constraint on the bulk surface $\Sigma$.\\
The question is considered in \cite{Castro:2014tta}, they conclude that the minimization of (\ref{3DCSformula}) results in the Mathisson-Papapetrou-Dixon(MPD) equations for a spinning particle in 3D, which is exactly the equation (\ref{3dConditionK}). To get the correction to HEE by CS term, one need to solve the MPD equation firstly\footnote{We would like to thank Prof. Takayanagi for reminding the paper \cite{Castro:2014tta} when preparing the draft.}.\\
Actually without knowing the equation of motion of the surface, one also could get the leading contribution of the correction to HEE. According to method of \cite{Schwimmer:2008yh}, the coordinate $y$ could be parameterized as
$y=\rho$ near the boundary, where $\rho$ is the coordinate of the bulk direction in the FG gauge, which states that any spacetime asymptotical to AdS admit the expansion
\begin{eqnarray}
ds^2= \frac{l^2}{4}\frac{d\rho^2}{\rho^2}+\frac{1}{\rho}(g_{(0)ij}+\rho g_{(1)ij}+\rho^2 g_{(2)ij}+...),
\end{eqnarray}
where $g_{(0)ij}$ is the boundary metric. Now we have two coordinates to describe the bulk metric, i.e., $X^{\mu}\in \{\rho,x_1,x_2\}$ and $\{z,\bar z,y\}$, where
$x_{1}$ and $x_{2}$ are the boundary coordinates. To find the leading contribution of the entropy one needs to know the transformation between the
two coordinates. It is possible to get the coordinate transformation near the boundary ($\rho \to 0$),as
\begin{eqnarray}\label{coordinatetrans}
&&\rho=y+B z y^{3/2}+ C \bar z y^{3/2}+...\nonumber \\
&&x^i= x^i_{0}(y)+A^iy^{1/2} z +\bar A^iy^{1/2} \bar z+...,
\end{eqnarray}
with $x^{i}_{0}=x^{i}_{0}(0)+D y...$, where $B,C,A^i,\bar A^{i},D$ are constants, there are some relation among these parameters, which is not important for our purpose.This transformation can be obtained by considering the following constraints.
\begin{eqnarray}
\frac{\partial X^{\mu}}{\partial z}\frac{\partial X^{\nu}}{\partial y}G_{\mu\nu}|_{z=0,\bar z=0}=0, \text{and} \frac{\partial X^{\mu}}{\partial z}\frac{\partial X^{\nu}}{\partial \bar z}G_{\mu\nu}|_{z=0,\bar z=0}=\frac{1}{2}.
\end{eqnarray}
We know from (\ref{3DCSformula}) that the additional terms for the HEE formula is proportional to
\begin{eqnarray}
S_{addition}\propto \int dy U_y.
\end{eqnarray}
$U_y$ can be written in the coordinate $\{\rho,x_1,x_2\}$ as
\begin{eqnarray}\label{ExpressionForU}
U_y\propto \frac{\partial X^{\mu}}{\partial y} \frac{\partial X^{\nu}}{\partial \bar z}\tri_{\mu} n_{z\nu},
\end{eqnarray}
where $n_{z}^{\nu} \equiv \frac{\partial x^{\nu}}{\partial z}$, the derivative $\tri$ is defined in the coordinate $\{\rho,x_1,x_2\}$. One could take
the coordinate transformation (\ref{coordinatetrans}) into (\ref{ExpressionForU}), and find the $\rho^{-1}$ term is vanishing, thus $U_y\propto O(\rho^0)$. As we know the first term in (\ref{3DCSformula}) would contribute a $\log$ divergence term for the HEE. So the additional term in the theory with CS term would not contribute to the leading
divergence. In \cite{Castro:2014tta} the authors calculate some examples in which the bulk are asymptotically to $AdS_3$, the result is consistent with conclusion above.

\section{Entropy of gravitational Chern-Simons terms in 7D}

We find the contribution from the 3D CS term for HEE or black hole entropy.  In 3D the possible correction related with extrinsic curvature do not appear.
It's also interesting to investigate this property in higher dimensional theory. It is well known that gravitational CS term only exists in ($4n-1$) dimensional
spacetime. We will discuss 7D theory in the following. Like the 3 dimensional case, one can't obtain the result directly by using
(\ref{Keyformula}). This is related with the fact that the action is not covariant. In higher dimension the trick that is used in \cite{Solodukhin:2005ah}
and \cite{Castro:2014tta} seems also difficult to operate. In this section we will use a ``topological method'' to get the result. We argue that the result is
correct. As an important check this method could produce the correct result for Wald entropy in arbitrary dimension. We will also discuss the trick that we have used in 3 dimension in section 5.

\subsection{Approach to 7D case by a topological method}

 In this section we would like to use a ``topological method'' to
derive the entropy for 7D theory with gravitational CS term. This method is based on the observation that the entropy of a topological Invariant is a local total derivative\footnote{After our paper we notice that a recent paper also uses the same method to deal with the entropy CS term\cite{Azeyanagi:2015uoa}. }. This is indeed the case for Euler densities, or equivalently, the Lovelock gravity in critical dimensions \cite{Dong:2013qoa}, see also \cite{Hung:2011xb}\cite{deBoer:2011wk}\cite{Guo:2013aca}. As we shall prove below, this is also the case for the Pontryagin density $tr(\bm{R}^4)$. Recall that we have
\begin{eqnarray}\label{7dChern-Simons}
d\Omega_{7}(\bm{\omega})=Tr\mathbf{R}^{4}.
\end{eqnarray}
We propose to derive the entropy of 7D CS term $s_7$ from the following identity
\begin{eqnarray}\label{7dCSentropy}
ds_7=\text{Entropy of}\  d\Omega_{7}(\bm{\omega})=\text{Entropy of} \ Tr\mathbf{R}^{4}.
\end{eqnarray}
Since the right hand side of (\ref{7dChern-Simons}) is invariant under the local Lorentz transformation, $s_7$ (up to an exact form) would also be free of the ambiguity. To make $s_7$ really be the entropy of 7d CS term, we have assumed
\begin{eqnarray}\label{7dCSentropy1}
\text{Entropy of}\  d\Omega_{7}(\bm{\omega})=d (\text{Entropy of} \ \Omega_{7}(\bm{\omega})).
\end{eqnarray}
An evidence for the above approach is that we derive the correct entropy with zero extrinsic curvature, which is obtained by using the generalized covariant phase formalism in \cite{Bonora:2011gz}\cite{Bonora:2012xv}.

Now let us start to derive the entropy of 7D CS term. We use the spin connection formulism in this section. Let's take a theory in 8D with the action,
\begin{eqnarray}\label{Action8D}
I_8=\int_{M_8}tr(\bm{R}^4).
\end{eqnarray}
By using the relation (\ref{Chern-Simons}) one have
\begin{eqnarray}
I_7=\int_{M_7} \bm{\Omega_7},
\end{eqnarray}
where $M_7$ is a 7D manifold as the boundary of $M_8$. The action (\ref{Action8D}) is invariant under the local Lorentz transformation. We would firstly get the entropy for the theory with such an action (\ref{Action8D}). The details of the calculation can be found in Appendix C, we list the result as follows.
\begin{eqnarray}\label{IndexForm}
S_8&=&-i\pi\int_{\Sigma_6}\sqrt{\det(g)}\hat \epsilon^{z\bar z i_1i_2i_3...i_6}\Big[-6K_{zi_1j_1}K_{\bar z i_2}^{\ \ j_1}R_{zj_2i_3i_4}R^{j_2}_{\ \bar z i_5i_6}\nonumber \\
&+&64\partial_{i_1}U_{i_2}\partial_{i_3}U_{i_4}K_{zi_5j_1}K_{\bar z i_6}^{\ \ j_1}+ 48K_{zi_1j_1}K_{\bar zi_2}^{\ \ j_1}\partial_{i_3}U_{i_4}K_{zi_5j_2}K_{\bar z i_6}^{\ \ j_2}\nonumber \\
&-&6K_{\bar z i_1}^{\ \ j_1}R_{zj_1i_2i_3}K_{zi_4j_2}R^{j_2}_{\ \bar z i_5 i_6}+2K_{zi_1j_1}r^{j_2j_1}_{\ \ i_2 i_3}K_{\bar z i_4j_3}r^{j_3}_{\ j_2 i_5 i_6}\nonumber \\
&-&12K_{zi_1j_1}r^{j_1}_{\ j_2i_2i_3}K_{\bar z i_4}^{\ \ j_2}K_{zi_5}^{\ \ j_3 }K_{\bar z j_3i_6}+8iK_{zi_1j_1}r^{j_1}_{\ j_2i_2i_3}K_{\bar z i_4}^{\ \ j_2}\partial_{i_5}U_{i_6}\nonumber \\
&+&64\partial_{i_1}U_{i_2}\partial_{i_3}U_{i_4}\partial_{i_5}U_{i_6}+8i\partial_{i_1}U_{i_2}R_{zj_1i_3i_4}R^{j_1}_{\bar z i_5i_6}\nonumber \\
&+&R_{zj_1i_1i_2}r^{j_1}_{\ j_2i_2i_3}R^{j_2}_{\ \bar z i_5i_6}\Big]+(z\leftrightarrow \bar z),
\end{eqnarray}
where the integration is on the codimension-2 surface $\Sigma_6$, ``-i'' appears because we are using Euclidean version. The result is still quite complex, to see it more clear, we would rewrite the result by forms. On the surface $\Sigma_6$, where $z=\bar z =0$, $K_{aij}$ and $U_i$ are one-form, $R_{aijk}$ and $r_{ijkl}$ are two forms.
One could map the other index into the local Lorenz coordinate by the vielbeins (\ref{vielbein}). For example\footnote{In the following,  $a$  , $a'$, etc, refers to $a_1,a_2$, $b$, $b'$, etc, refers to $b_1,b_2,...$ , c refers to $a_1,a_2,a_3...$},
\begin{eqnarray}
K_{zij}= e_z^{a}e_i^b\omega_{i,ba},\ \ -2iU_i = e^{z a'}e_{z}^{a''}\omega_{i,a' a''},\ \ r_{ijkl}=e^{b'}_ie^{b''}_jr_{b'b''kl},
\end{eqnarray}
one could rewrite (\ref{IndexForm}) as\footnote{We note that $dx^{\nu_1}\land dx^{\nu_2}...\land dx^{\nu_n}=\epsilon^{\nu_1 \nu_2...\nu_n}\sqrt{\det(g)}d^nx$.}
\begin{eqnarray}
S_8&=&-i\pi\int_{\Sigma_6}\Big[-6\epsilon^{aa'}\bo_{ba}\bo^b_{\ a'} \bR_{a''b'}\bR^{b'a''}+64d\bm{U}d\bm{U}\epsilon^{aa'}\bo_{ba}\bo^b_{\ a'}\nonumber \\
&+&24d\bm{U}(\epsilon^{aa'}\bo_{ba}\bo^b_{\ a'})^2+8\bo_{ba}\bR^{ba}\epsilon^{a'a''}\bo_{b'a'}\bR^{b'}_{\ a''}\nonumber \\
&+&8\epsilon^{aa'}\bo_{ba}\bm{r}^{b}_{\ b'}\bm{r}^{b'}_{\ b''}\bo^{b''}_{\ a'}-6\bo_{ba}\bm{r}^{b}_{\ b'}\bo^{b'a}\epsilon^{a'a''}\bo_{b''a'}\bo^{b''}_{\ a''}\nonumber \\
&+&32\bo_{ba}\bm{r}^{b}_{\ b'}\bo^{b'a}d\bm{U}+128d\bm{U}d\bm{U}d\bm{U}+32d\bm{U}\bR_{ab}R^{ba}\nonumber \\
&+&8\epsilon^{aa'}\bR_{ab}\bm{r}^{b}_{\ b'}\bR^{b'}_{\ a'}.
\end{eqnarray}
To simplify result we need the relation
\begin{eqnarray}
\bR_{ab}\equiv d\bo_{ab}+\bo_{ac}\bo^c_{\ b}=d\bo_{ab}+\bo_{aa'}\bo^{a'}_{\ b}+\bo_{ab'}\bo^{b'}_{\ b},
\end{eqnarray}
\begin{eqnarray}
\bm{r}_{bb'}=d\bo_{bb'}+\bo_{bb''}\bo^{b''}_{\ b'}.
\end{eqnarray}
The result is
\begin{eqnarray}\label{Interestingresult}
S_8=\int_{\Sigma_6}d\bm{s}_{7},
\end{eqnarray}
with
\begin{eqnarray}
\bm{s}_7&=&-i\pi\Big[8\epsilon^{aa''}\bo_{ab}\bo^b_{\ b'}\bo^{b'}_{b''}\bo^{b''}_{\ c}\bo^{c}_{\ a''}-6\epsilon^{aa'}\bo_{ba}\bo^{b}_{\ a'}\bo_{a''b''}R^{b''a''}\nonumber \\
&+&64\bm{U}d\bm{U}\epsilon^{aa'}\bo_{ba}\bo^b_{\ a'}+32\bm{U}d\bo_{ab}d\bo^{ba}+128\bm{U}d\bm{U}d\bm{U}\nonumber \\
&+&8\epsilon^{aa'}\bo_{ab}d\bo^{b}_{\ b'}\bo^{b'}_{a''}\bo^{a''}_{a'}-8\epsilon^{aa'}\bo_{ba}\bo^{b}_{\ b'}d\bo^{b'}_{\ b''}\bo^{b''}_{\ a'}\nonumber \\
&+&\epsilon^{aa'}d\bo_{ba}\bo^{b}_{\ b'}\bo^{b'}_{\ b''}\bo^{b''}_{\ a'}+32d\bm{U}\bo^{aa'}\bo_{ba}\bo^b_{\ a'}\nonumber \\
&+&16\bm{U}\bo_{ab}d\bo^{b}_{\ b'}\bo^{b'a}-32d\bm{U}\bo_{ab}\bo^{b}_{\ b'}\bo^{b'a}\nonumber \\
&+&8\epsilon^{aa'}\bo_{ab}d\bo^{b}_{\ b'}d\bo^{b'}_{a'}
\Big].
\end{eqnarray}
One could define the ``density'' of the entropy in 8D $\bm{s}_8$, and the relation $\bm{s}_8=d \bm{s}_7$. This is the expected relation that we mentioned in the beginning of this subsection. $\bm{s}_8$ is the entropy that we obtain from the Pontryagin density, it has the similar relation (\ref{Chern-Simons}) which must be satisfied by the Pontryagin class.
The entropy for the CS term in 7D is
\begin{eqnarray}
S_7=\pi\int_{\Sigma_5}(\bm{s}_7+d\bm{s'}),
\end{eqnarray}
where the surface $\Sigma_5$ is the codimension-2 surface in 7D, which is also a boundary of $\Sigma_6$\footnote{Actually the action in 8D depends on one more coordinate, the result of the corresponding entropy is dependent on the coordinate. But one can assume to choose a suitable Mainfold for $M_8$, on which the boundary is $M_7$, and the boundary of the surface $\Sigma_6$ is $\Sigma_5$ }, $\bm{s'}$ is arbitrary. Our above approach actually do not use the viebeins in the regularized spacetime (\ref{vielbein}), it is not expected the result is effected by the local gauge transformation. \\

Here we only calculate the 7D result. We expect it can be generalized to $(4n+1)$D without any difficulty in principle. Conversely, we could say
our result above provides another evidence to support our proposal that the entropy of a topological Invariant is also a topological Invariant.

\section{ Wald entropy in arbitrary dimension}

\subsection{The topological method}
Here we use a “topological approach” to this problem by considering the relation (\ref{Chern-Simons}). For 3 dimension
\begin{eqnarray}
d\bm{\Omega_3}=Tr \bm{R}^2,
\end{eqnarray}
here $\bm{R}$ lives in 4 dimension spacetime $M_4$, $\bm{\Omega}_3$ lives in $M_3$. The above terms are invariant under local Lorentz transformation. Let's define
\begin{eqnarray}\label{4R}\label{I4}
\tilde{I}_4=\int_{M_4} Tr \bm{R}^2=\int_{M_3} \bm{\Omega}_3.
\end{eqnarray}
It's easy to check the contribution from $(\partial_z A,\partial_{\bar z}A)$ vanishes. The total contribution of (\ref{4R}) to the entropy is
\begin{eqnarray}\label{S_4}
\tilde{S}_4= 8i\pi \int_{\Sigma_2} \bm{R_N},
\end{eqnarray}
where the integration is on the a dimension-2 surface $\Sigma_2$, $\bm{R_N}$ is defined as
\begin{eqnarray}
\bm{R_N}\equiv \frac{1}{2}tr{\epsilon \bm{R}}.
\end{eqnarray}
Note that $\bm{R_N}=d \bm {\Gamma_N}$, where ``$d$'' is the defined on $\Sigma_2$,
\begin{eqnarray}
\bm{\Gamma_N}\equiv \frac{1}{2}tr(\epsilon \bm {\Gamma}).
\end{eqnarray}
One could rewrite (\ref{S_4}) as
\begin{eqnarray}
\tilde{S}_4=8i\pi\int_{\Sigma_1} \bm{\Gamma_N},
\end{eqnarray}
where $\Sigma_1$ is a codimension-3 surface in $M_4$, as well as a codimension-2 surface in $M_3$. Formally considering (\ref{I4}) we have the entropy
\begin{eqnarray}
S_3=8i\pi\int_{\Sigma_1} \bm{\Gamma_N},
\end{eqnarray}
if one uses the trick to find suitable $M_4$ such that $M_3$ is a boundary of $M_4$, and  $\bm{\Sigma}_1$ as a boundary of $\bm{\Sigma}_2$.  \\
The result is same as (\ref{3DCSformula}), and also \cite{Castro:2014tta}\cite{Solodukhin:2005ah}.
Now it is easy to generalize the method to higher dimension. Actually the generalization is quite trivial.
With the relation
\begin{eqnarray}
d\bm{\Omega}_{2n+1}=tr \bm{R}^{n+1},
\end{eqnarray}
one could have
\begin{eqnarray}
\tilde{I}_{2n+2}= \int_{M_{2n+2}} tr \bm{R}^{n+1}=\int_{M_{2n+1}}\bm{\Omega}_{2n+1},
\end{eqnarray}
The entropy from $\tilde{I}_n$ is
\begin{eqnarray}
\tilde{S}_{2n+2}=4i\pi (n+1)\int_{\Sigma_{2n}}\bm{R_N}^{n}.
\end{eqnarray}
With the relation $\bm{R_N}=d \bm {\Gamma_N}$, one have
\begin{eqnarray}\label{FormalmethodWaldentropy}
S_{2n+1}=4i\pi (n+1)\int_{\Sigma_{2n-1}}\bm{\Gamma_N}\bm{R_N}^{n-1}.
\end{eqnarray}
The result is consistent with the one that is obtained by the covariant phase formalism \cite{Bonora:2011gz}\cite{Bonora:2012xv}.
\subsection{Another approach}
Actually we also could use the trick in 3 dimension to get the correct result for Wald entropy in arbitrary dimension. We will briefly state the process in the following. As the appendix \ref{Sectionnovariant} shows, a local gauge transformation would lead to a boundary term which is related with $\epsilon$ in the regularized spacetime. Even if we only consider the Wald entropy,i.e., the extrinsic curvature is vanishing, this ambiguity still exist. The following term is suitable to add to the action\footnote{The CS action is normalized as
\begin{eqnarray}
I_{n}=(n+1)\int_M \bm{\Omega_{2n+1}}.
\end{eqnarray}
},
\begin{eqnarray}\label{AdditionalTerm}
\Delta I_{n}=-n(n+1)\int_0^1 dt(t-1) t^{n-1}\int_M d str(\bm{\omega}(0),\bm{\omega}(\epsilon), \bm{R_t}^{n-1}),
\end{eqnarray}
where $\bm{\omega}(\epsilon)$ is the spin connection in the regularized metric (\ref{metric}), $\bm{R_t}$ is constructed by $\bm{\omega}(\epsilon)$.
After complex calculation one could get the variance of the action with the additional term,
\begin{eqnarray}\label{Variancetotalaction}
 \delta_\theta (I_n+\Delta I_n)&=& -n(n+1)\int_0^1 dt (t-1) t^{n-1}\int_M\Big[-str(\bm{R_1},d\theta, \bm{R_t}^{n-1})  \nonumber \\
&+&str(\bm{\omega}(0),\bm{R_2},\bm{R_t}^{n-1})\Big]\nonumber \\
&+& 2n(n+1)\int_0^1 dt (t-1)^2 t^{n-1}\int_M \Big[str(\bm{R_1},\bm{\omega},\bm{R_2},\bm{R_t}^{n-2})\nonumber \\
&-&str(\bm{\omega}(0),D_t\bm{\omega},\bm{R_2},\bm{R_t}^{n-2})+str(\bm{\omega}(0),\bm{\omega},\bm{R_3},\bm{R_t}^{n-2})\Big],
\end{eqnarray}
with the following definitions,
\begin{eqnarray}\label{Newdefinedcur}
\bm{R_1}=D_t \bm{\omega}(0),\ \ \ \bm{R_2}=D_t d\theta,\ \ \ \bm{R_3}=D_t D_t d\theta=[\bm{R_t},d\theta],
\end{eqnarray}
see the definition of ``$D_t$'' in Appendix B. One could extract the $O(\epsilon)$ terms in (\ref{Variancetotalaction}), and finally get
\begin{eqnarray}\label{EntropyGaugetransformation}
\Delta S= -4i\pi(n+1)\int_\Sigma df \bm{R_N}^{n-1},
\end{eqnarray}
where $\Sigma$ is the codimension-2 surface, $f$ is defined as
\begin{eqnarray}\label{EntropyGaugetransformationf}
f\equiv \frac{n-1}{n+1}tr(\theta \epsilon).
\end{eqnarray}
$R_N$ is defined as
\begin{eqnarray}\label{RN}
\bm{R_N}\equiv \frac{1}{2}tr(\epsilon \bm{R})
\end{eqnarray}
where $\epsilon$ is 2-dimension Levi-Civita symbol with $\epsilon_{12}=-\epsilon_{21}=1$. (\ref{EntropyGaugetransformationf}) is actually a integration of total derivative term. Thus the local gauge transformation will not effect the final result of the entropy, which is what we expect. With the additional term one could get the entropy,
\begin{eqnarray}\label{Waldentropy}
S=4i\pi(n+1) \int_\Sigma \bm{\omega}_N \bm{R_N}^{n-1},
\end{eqnarray}
where $\bm{R_N}$ is defined by (\ref{RN}), and $\bm{\omega}_N$ is defined as
\begin{eqnarray}\label{omegaN}
\bm{\omega}_N=\frac{1}{2}tr(\epsilon \bm{\omega}),
\end{eqnarray}
the result is consistent with \cite{Bonora:2011gz}\cite{Bonora:2012xv} and (\ref{FormalmethodWaldentropy}).
(\ref{RN}) and (\ref{omegaN}) are related with each other by
\begin{eqnarray}
d \bm{\omega_N}=\bm{R_N},
\end{eqnarray}
for $tr(\epsilon \bm{\omega}\bm{\omega})=0$. Then it's obvious (\ref{EntropyGaugetransformation}) can be seen as the
gauge transformation
\begin{eqnarray}
 \bm{\omega_N}\to \bm{\omega_N}-df.
 \end{eqnarray}
Note that in 3D $f=0$, so the result in 3D is gauge invariant.\\
We use the spin connection formulism of CS action in the above discussion. But it's easy to generalize the result to  Christoffel symbols formulism. The non-covariant part of the diffeomorphism $\delta_{\xi}$ of $\bm{\Gamma}$ is,
\begin{eqnarray}
\hat \delta_{\xi}\bm{\Gamma}=d \Lambda,
\end{eqnarray}
where $\Lambda^a_{\ b}=\partial_b \xi^b$. If one replaces the parameter $\theta$ with $\Lambda$,  as well as $\bm{\Gamma}$ with $\bm{\omega}$, all the result for Christoffel symbols formulism can be obtained.\\

\section{Conclusion and Discussion}
In this paper we have analyzed the entropy when gravitational Chern-Simons terms are added into the action.  We find it is necessary to modify the squashed cone method for Chern-Simons theory. The necessity is related with the non-covariant part transformation of diffeomorphism for a total derivative term would appear under such transformation. The covariant theory is free of this ambiguity. One possible solution to this
problem is to add a total derivative term into the original Lagrangian, which does not affect the equation of motion, at the same time eliminates the ambiguity caused by the diffeomorphism. This term is also vanishing in the limit $\epsilon \to 0$.\\
For gravitational Chern-Simons term we suggest a term (\ref{AdditionalTerm}) which could lead to a consistent result for the special case $K_{aij}=0$. We find the gauge freedom of the Chern-Simons action is not completely broken by such term, there is still a gauge transformation on the codimension-2 subspace for the entropy. This may lead us to find a principle to fix the special term. For the general case, i.e., $K_{aij}\ne 0$, we suspect that (\ref{AdditionalTerm}) is sufficient to get a consistent result. It seem more terms are needed to eliminate the ambiguity of the anomaly entropy.  It's worthy to go on the discussion on this direction. On the other hand we propose a `topological approach' to calculate the entropy of gravitational Chern-Simons terms when the extrinsic curvature $K_{aij}$ is non-vanishing. It yields the correct Wald entropy in arbitrary dimension and gives non-trivial results when the extrinsic curvature does not vanish. Our results imply that the entropy of a topological invariant seems to also be a topological invariant. There may exist some mathematical interpretations or correspondence for this nice property. We hope someone could clarify this problem in future.

\section*{Acknowledgements}

R. X. Miao is supported by Sino-Germann (CSC-DAAD) Postdoc Scholarship Program. W. Z. Guo is supported by Postgraduate Scholarship Program of China Scholarship Council. We thank Prof. Miao Li for his encouragement and support.
We are grateful to School of Astronomy and Space Science at Sun Yat-Sen University for hospitality where part of work was done. We thank Tadashi Takayanagi, Song He, Noburo Shiba for discussions.

\appendix
\section{Useful components of $\bm{\Gamma}$ and $\bm{R}$}
\begin{eqnarray}
\Gamma_{\bar z\bar z}^{\bar z}=2\partial_{\bar z}A,
\ \ \ \Gamma_{i\bar z}^{\bar z}=2iU_{i},
\ \ \ \Gamma^j_{\bar z i}=g^{jk}K_{\bar z ki},
\ \ \ \Gamma_{ij}^z=-2e^{-2A}K_{\bar zji},
\ \ \Gamma_{li}^m =\gamma_{li}^m,
\end{eqnarray}
\begin{eqnarray}\label{spinc1}
&&\omega_{z\ \ z}^{\ z}=-\omega_{z\ \ \bar{z}}^{\ \bar{z}}=\partial_z A,\ \
\omega_{i\ \ z}^{\ z}=-2i U_i,\ \ \omega_{i\ \ z}^{\ j}=K_{z\ i}^{\ j},\ \ \omega_{i\ \ j}^{\ z}=-2e^{-2A}K_{\bar{z}\ i}^{\ j},\ \ \omega_{i\ \ j}^{\ k}=\bar{\omega}_{i\ \ j}^{\ k},
\end{eqnarray}

\begin{eqnarray}
&&R_{z \bar z z\bar z}=e^{2A}\Big(\partial_z \partial_{\bar z}A-3Te^{2A} \Big),\nonumber \\
&&R_{z\bar z z i}=\frac{1}{2}e^{2A}\Big[2iU_{i}(\bar z \partial_z \partial_{\bar z}A+\partial_z A +z\partial_z \partial_z
A)+3iV_{zi}\Big]\nonumber \\
&&R_{zizj}=\frac{1}{2}\Big[4K_{zij}\partial_z A-2Q_{zzij}+2g^{lk}K_{zlj}K_{zik}\Big]\nonumber \\
&&R_{z\bar zij}=\frac{1}{2}e^{2A}\Big[2i\partial_i U_j-2e^{-2A}g^{ln}K_{\bar znj}K_{zli}\Big]-(i\leftrightarrow j),\nonumber \\
&&R_{z i \bar z j}=\frac{1}{2}e^{2A}\Big[i(\partial_i U_j -\partial_j U_i)+2e^{-2A}g^{mn}K_{\bar z ni}K_{zmj}+4U_i U_j-2e^{-2A}Q_{z\bar zij}\Big]\nonumber \\
&&R_{zijk}=\frac{1}{2}e^{2A}\Big[-2e^{-2A}\partial_j K_{zki}-4ie^{-2A}U_jK_{zik}-2e^{-2A}K_{zlj}\gamma_{ik}^l\Big]-(j\leftrightarrow k)\nonumber \\
&&R_{ikjl}=r_{ikjl}-\Big[2e^{-2A}(K_{zij}K_{\bar z kl}+K_{\bar z ij}K_{zkl})\Big]-(j\leftrightarrow k).
\end{eqnarray}

\section{Non-covariant part of $\bm{\Omega}$}\label{Sectionnovariant}
 Following the step as \cite{Bonora:2011gz}, we define the ``covariant'' derivative
\begin{eqnarray}
D\  =d\  +[\bm{\omega},\  ],\quad \quad \quad D_t\ =d\ +[t\bm{\omega},\ ],
\end{eqnarray}
note that $\frac{d}{dt}(t\bm{R_t})= D_t \bm{\omega}$
one can get
\begin{eqnarray}
\hat{\delta}\bm{\Omega}_{2n+1}&=&(n+1)\int_0^1 dt t^n\Big( str( \hat{\delta} \bm{\omega},\bm{R_t}^n)+nstr( \bm{\omega},\hat{\delta} \bm{R_t},\bm{R_t}^{n-1})\Big)\nonumber \\
&=&(n+1)\int_0^1 dt\Big(t^n str( d\theta,\bm{R_t}^n)+n t^{n-1} str( \bm{\omega},(t-1)D_t d\theta,\bm{R_t}^{n-1})\Big)\nonumber \\
&=&(n+1)\int_0^1 dt\Big(t^n str( d\theta,\bm{R_t}^n)-n t^{n-1}d str( \bm{\omega},(t-1) d\theta,\bm{R_t}^{n-1})\nonumber \\
&+& n str(\frac{d}{dt}(t\bm{R_t}),(t-1)d\theta, (t\bm{R_t})^{n-1}\Big),
\end{eqnarray}
where the last step we use
\begin{eqnarray}
 d str(\bm{A_1},\bm{A_2}...,\bm{A_n})=\sum_{i=1}^n(-)^{a_1+a_2...+a_{i}}str(\bm{A_1},...,D(\bm{A_i}),...),
\end{eqnarray}
$\bm{A_i}$ denotes the $a_i$-form, the covariant derivative $D\ \equiv d\ +[\Theta,\ ]$, $\Theta$ is 1-form.
Then
\begin{eqnarray}
&&\int_0^1 dt n str(\frac{d}{dt}(t\bm{R_t}),(t-1)d\theta, (t\bm{R_t})^{n-1}=\nonumber \\
&&\int_0^1 dt\Big( \frac{d}{dt} str((t-1)d\theta,t^n\bm{R_t}^{n})- t^n str( d\theta,\bm{R_t}^n) \Big),
\end{eqnarray}
the first term vanishes. We get
\begin{eqnarray}\label{varianceofCSaction}
\hat{\delta}\bm{\Omega}_{2n+1}=-n(n+1)d\int_0^1 dt t^{n-1} (t-1) str( \bm{\omega},d\theta,\bm{R_t}^{n-1}).
\end{eqnarray}

\section{Details of the calculation in section 5}
The action (\ref{Action8D}) can be  written by the spacetime components as
\begin{eqnarray}
I_8=\frac{1}{16}\int_{M_8}\sqrt{det(G)}\epsilon^{\beta_1\beta_2...\beta_8}R^{\alpha_1}_{\ \alpha_2 \beta_1\beta_2}R^{\alpha_2}_{\ \alpha_3 \beta_3\beta_4}R^{\alpha_3}_{\ \alpha_4 \beta_5\beta_6}R^{\alpha_4}_{\ \alpha_1 \beta_7\beta_8}.
\end{eqnarray}
The contribution to the entropy from $R_{z\bar z z\bar z}$ is
\begin{eqnarray}
I^{(1)}&=&2\int_{M_8}\sqrt{det(G)}\epsilon^{z\bar z i_1i_2i_3i_4i_5i_6}R_{z\bar zz\bar z}\Big[4R_{z\bar z i_1i_2}R_{z\bar zi_3i_4}R_{z\bar zi_5i_6}\nonumber \\
&+&2R_{z\bar zi_1i_2}R_{zj_1i_3i_4}R^{j_1}_{\ \bar z i_5i_6}+2R_{zj_1i_1i_2}R^{j_1}_{\ \bar z i_3i_4}R_{z\bar z i_5i_6}\nonumber \\
&+&R_{zj_1i_1i_2}R^{j_1}_{\ j_2i_3i_4}R^{j_2}_{\ \bar zi_5i_6}
\end{eqnarray}
The contribution from $(R_{zizj}, R_{\bar zi \bar z j})$ is
\begin{eqnarray}
I^{(2)}&=&4\int_{M_8}\sqrt{det(G)}\epsilon^{z\bar z i_1i_2i_3i_4i_5i_6}\partial_zA \partial_{\bar z}A\Big[R_{zj_1i_1i_2}R^{j_1}_{\ \bar zi_3i_4}K_{zi_5j_2}K_{\bar z i_6}^{\ \ j_2}(G^{z\bar z})^2\nonumber \\
&+&K_{zi_1j_1}K_{\bar z i_2}^{\ \ j_2}R_{z\bar zi_3i_4}R_{z\bar z i_5i_6}(G^{z\bar z})^3+K_{zi_1}^{\ \ j_1}K_{\bar z i_2j_2}R^{j_2}_{\ zi_3i_4}R_{\bar zj_1i_5i_6}(G^{z\bar z})^2\nonumber \\
&+&K_{zi_1}^{\ \ j_1}K_{\bar z i_2j_2}R^{j_2}_{\ \bar zi_3i_4}R_{ zj_1i_5i_6}(G^{z\bar z})^2+K_{zi_1}^{\ \ j_1}K_{\bar zi_1 j_2}R^{j_3}_{\ j_4i_3i_4}R^{j_3}_{\ j_1i_5i_6}G^{z\bar z}\nonumber \\
&-&K_{zi_1j_1}R^{j_1}_{\ zi_2i_3}K_{\bar z i_4j_2}R^{j_2}_{\ \bar z i_5i_6}+K_{zi_1j_1}R^{j_1}_{\ j_2i_2i_3}K_{\bar z i_4}^{\ \ j_2}R_{z\bar zi_5i_6}\Big]+(z\leftrightarrow \bar z).
\end{eqnarray}
One could get the $O(\epsilon)$ term of the above terms after complex calculation,

\begin{eqnarray}
S_8&=&\pi\int_{\Sigma_6}\sqrt{\det(g)}\hat \epsilon^{z\bar z i_1i_2i_3...i_6}\Big[-6K_{zi_1j_1}K_{\bar z i_2}^{\ \ j_1}R_{zj_2i_3i_4}R^{j_2}_{\bar z i_5i_6}\nonumber \\
&+&64\partial_{i_1}U_{i_2}\partial_{i_3}U_{i_4}K_{zi_5j_1}K_{\bar z i_6 j_1}+ 48K_{zi_1j_1}K_{\bar zi_2}^{\ \ j_1}\partial_{i_3}U_{i_4}K_{zi_5j_2}K_{\bar z i_6 j_2}\nonumber \\
&-&6K_{\bar z i_1}^{\ \ j_1}R_{zj_1i_2i_3}K_{zi_4j_2}R^{j_2}_{\ \bar z i_5 i_6}+2K_{zi_1j_1}r^{j_2j_1}_{\ \ i_2 i_3}K_{\bar z i_4j_3}r^{j_3}_{\ j_2 i_5 i_6}\nonumber \\
&-&12K_{zi_1j_1}r^{j_1}_{\ j_2i_2i_3}K_{\bar z i_4}^{\ \ j_2}K_{zi_5}^{\ \ j_3 }K_{\bar z j_3i_6}+8iK_{zi_1j_1}r^{j_1}_{\ j_2i_2i_3}K_{\bar z i_4}^{\ \ j_2}\partial_{i_5}U_{i_6}\nonumber \\
&-&32i\partial_{i_1}U_{i_2}\partial_{i_3}U_{i_4}\partial_{i_5}U_{i_6}+8i\partial_{i_1}U_{i_2}R_{zj_1i_3i_4}R^{j_1}_{\bar z i_5i_6}\nonumber \\
&+&R_{zj_1i_1i_2}r^{j_1}_{\ j_2i_2i_3}R^{j_2}_{\ \bar z i_5i_6}\Big]+(z\leftrightarrow \bar z).
\end{eqnarray}

\end{document}